\documentclass[twocolumn,epsf,prd,floatfix,nofootinbib]{revtex4}
\usepackage{graphicx}
\usepackage{epsfig}
\usepackage{dcolumn}
\usepackage{bm}
\usepackage{latexsym}

\begin{document}

\title{Polarization of $\tau$ leptons produced in ultra-high energy neutrino-nucleon scattering}
\author{K\'evin Payet$^1$}

\affiliation{
$^1$ LPSC, Universit\'e Joseph Fourier Grenoble 1, CNRS/IN2P3, INPG, Grenoble, France\\}

\begin{abstract}
We study the polarization vector of $\tau^-$ ( $\tau^+$ ) leptons created in $\nu_\tau$ ($\overline{\nu}_\tau$) - Nucleon deep inelastic scattering. Our work is oriented toward ultra-high
energy particles relevant for cosmic-rays experiments. The goal of this paper is to derive relevant informations to constrain the systematic uncertainties that follow from the
lack of knowledge about this point particularly when studying Earth-skimming neutrinos.
\end{abstract}

\maketitle

\section{Introduction}

Through the last years, the observation of ultra-high energy (UHE) neutrinos has become one of the challenges of astroparticle physics. Many models, either astrophysical or exotic
models, predict a substantial flux of neutrinos. One of the most certain contribution to this neutrino flux are the so-called GZK-neutrinos produced in the decay of pions and
kaons, from the interaction of ultra-high energy protons with the CMB \cite{GZK}.

Such a mechanism provides a substantial flux of muon and electron neutrinos at the point of interaction. But given the large distances traveled by the particles, an observer can expect
equal fluxes of $\nu_e$, $\nu_\mu$ and $\nu_\tau$ at the observation point, due to flavour mixing and oscillations \cite{osc1,osc2}.

During the last years, an increasing effort has been put forward to develop a new generation of dedicated neutrino telescopes, both in the southern \cite{Andres:1999hm,ice3} and
the northen \cite{antares,nemo,nestor,Aynutdinov:2005sc} hemisphere.
Such detectors are relevant for an energy range of $10^{-6}$ EeV $\leq E_\nu \leq 10^{-1}$ EeV. At higher energies other experiments develop the detection of coherent radio
emission produced by neutrino-induced showers in matter \cite{anita,rice,forte}.
Ultra-high energy cosmic-ray detectors such as the Pierre Auger Observatory \cite{augergen} and the HiRes Fly's Eye detector \cite{hires}, although they were not developed for the detection of
neutrinos, may have equal or even better potential in the ultra-high energy range of $10^{-1}$ EeV $\leq E_\nu \leq 10^2$ EeV, where the GZK-neutrinos are expected \cite{cosmogenic}.

In addition to the classical way of detecting neutrinos, at large zenith angles ($\theta > 75^\circ$) \cite{beresmirnov,capelle}, it has been pointed out recently that the detection potential could
be enhanced by the presence of $\nu_\tau$, due to oscillations, in the cosmic neutrinos flux \cite{antoine,fargion}. Upward-going UHE tau neutrinos that graze the Earth just below the horizon (also
called "Earth-skimming neutrinos") have a quite high probability to interact in the crust and produce a tau lepton which, if produced close enough from the surface, may emerge
and trigger an extensive air shower which may be detected by a surface detector, provided it does not decay too far from the ground.

The estimation of the sensitivity to such UHE neutrinos implies the use of Monte Carlo simulations in which the propagation of tau neutrinos and tau leptons, produced in
$\nu_\tau$ charged current (CC) interactions, is simulated. To obtain the best description, all relevant processes for the energy scale considered must be taken into account.
The ultra-high energy cosmic ray range dealt with here implies large extrapolations for the physical quantities considered in the calculation such as the parton distribution
functions (pdf) involved in the simulation of the interactions with the nucleons of the crust. These large extrapolations lead to large systematic uncertainties which must be
studied precisely to estimate the relevance of the different results obtained from these Monte Carlo calculations. In the estimate of the sensitivity to
Earth-skimming neutrinos, the uncertainties follow for a big part from the CC and NC deep inelastic cross-sections (related to the pdf uncertainties) and from the tau lepton 
energy loss along its propagation. Furthermore, during $\nu_\tau$ CC interaction, the tau leptons created are likely to be highly polarized. This polarization plays also an
important role in estimating the systematics as it can impact, mainly on the decay of the $\tau$ lepton, and in a smaller extent on its propagation. When a
tau decays, the energy distribution of the different products depends drastically on the helicity of the decaying tau lepton \cite{tauola} and
this affects directly the
development of extensive air showers (EAS) triggered by such decays, thereby influencing the detection and identification of such air showers. For surface detectors based on the 
study of EAS, the tau polarization can thus be one of the largest source of systematic uncertainties, through the calculation of the acceptance \cite{pao1}.

The aim of this paper is to study more precisely the polarization of tau leptons produced in ultra-high energy $\nu_\tau$ CC interactions. Our goal is to constrain as much
as possible this source of systematic uncertainties for experiments that are sensitive to the decay products of the tau lepton. We use a Monte Carlo calculation
in which CC interactions of UHE tau neutrinos are simulated and the polarization of the induced tau lepton is derived for each interaction. In section \ref{sec:formula}, we 
present the different ingredients used to derive the tau polarization. In section \ref{sec:results}, we apply this framework to mono-energetic beams of tau neutrinos and present the 
results obtained.Finally, we discuss our conclusions in section \ref{sec:conclusion}.

\section{Calculation of tau polarization}
\label{sec:formula}

We study the tau polarization through a Monte Carlo procedure. We simulate a neutrino beam that is forced to interact through CC interaction and for each interaction we
compute the spin polarization of the produced tau.

The reaction to be described is the following: $\nu_\tau (p_\nu) N (p) \rightarrow \tau (p_\tau) X$, where $q = p_\nu - p_\tau$ is the transfered 4-momentum. 
\begin{figure}[t]
    \centering
    \includegraphics[width=6.66cm,height=4cm]{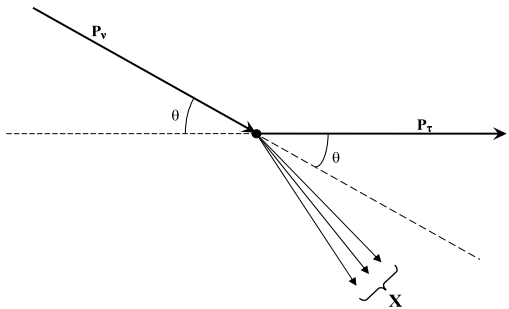}
    \caption{\small{Scheme of a $\nu_\tau$-nucleon deep inelastic interaction in the laboratory frame.}}
    \label{fig:shema}
\end{figure}
The $x$ axis is defined along the direction of propagation of the $\tau$ lepton. The y axis is chosen so that the scattering plane corresponds to the $x-y$ plane.

In all the following we consider an isoscalar nucleon target.

\subsection{Ingredients}
\label{subsec:ingredients}

For the computation of the polarization, we followed the work of Levi \cite{levi} who derived practical formulae for the polarization vector of the tau lepton produced in deep
inelastic neutrino-nucleon scattering. Once we have a simple formula to compute what we want to study (see equations \ref{eq:Ptau}, \ref{eq:Ptaubar} and \ref{pqs}), we still need some other ingredients. First, simulating a CC
deep-inelastic interaction is nothing more than drawing a pair of bjorken variables $(x,y)$ that fully characterise our interaction. In the deep inelastic frame, these
dimensionless variables
are defined as follow

\begin{equation} \begin{array}{l} y=\frac{E_\nu-E_\tau}{E_\nu} \\ \\ x=\frac{Q^2}{2ME_\nu y}
		  \end{array}
\end{equation}
These definitions are expressed here in the laboratory frame (fixed nucleon target) for simplicity but their values are frame independent. $E_\nu$ is the incident neutrino energy, $E_\tau$ the
energy of the produced tau and $Q^2 = -q^2$. M is the nucleon mass. We thus need an $(x,y)$ distribution to simulate the interaction.

For the computation of these distributions as well as that of the tau polarization, we also need to use a pdf set to evaluate the distribution of partons inside the nucleon. Two
different sets were used to
check the dependency of our results upon this choice. We chose CTEQ6 \cite{CTEQ} and MRST distributions \cite{MRST}, that are both fitted to experimental data, where available.

For the high energies considered in our work, regions of very small $x$ can be probed during the scattering (see equation \ref{eq:xlimit}), where no experimental informations
exist. We thus have to control the extrapolation 
to these small $x$ in a secure way. For that we used the prescription of Reno \cite{reno}:

\begin{equation}
x\overline{q}(x,Q^2) = \left(\frac{x_{min}}{x}\right)^\lambda x\overline{q}(x_{min},Q^2),\mbox{ \ \ \ for $x < x_{min}$}
\label{eq:xextrapolation}
\end{equation}

We chose $x_{min}=10^{-6}$, which is the lower limit for CTEQ6 set. $\lambda$ is determined for each flavour from the fit of the density function with $Q=M_W$.

We derived our own distributions in $x$ and $y$ from the expression of the differential cross-section given in reference \cite{gandhi}:

\begin{eqnarray} 
\frac{d^2\sigma}{dxdy} = \frac{2G^2_FME_\nu}{\pi}\left(\frac{M^2_W}{Q^2+M^2_W}\right)^2\times
\nonumber\\
\times[x\mathcal{Q}(x,Q^2)+x\overline{\mathcal{Q}}(x,Q^2)(1-y^2)],  
\label{eq:diffXsect}
\end{eqnarray}

$M_W$ is the mass of the W boson and $G_F=1.16632\times 10^{-5} $ GeV$^{-2}$ is Fermi's constant.

$\mathcal{Q}$ and $\overline{\mathcal{Q}}$ are the quarks distribution functions and are written from the individual distributions of valence and sea quarks obtained from the
pdf sets mentionned above. For exact expressions see the previous reference. 

The formula for  $\overline{\nu}_\tau$ scattering is obtained from equation \ref{eq:diffXsect} mostly by interchanging $\mathcal{Q}$ and $\overline{\mathcal{Q}}$, though it must be noted
that the exact expressions for these two distribution functions are not exactly the same for $\nu_\tau$ and $\overline{\nu}_\tau$.

The physical regions for $x$ and $y$ are obtained by Albright and Jarlskog \cite{albright}, with $m_\tau$ the mass of the $\tau$:

\begin{equation} \frac{m^2_\tau}{2M(E_\nu-m_\tau)} \leq x < 1 \label{eq:xlimit}\end{equation}
and
\begin{equation} A-B \leq y \leq A+B
\label{eq:yrange}
\end{equation}
where
\begin{equation} \begin{array}{l} A=\left.\frac{1}{2} \left( 1-\frac{m^2_\tau}{2ME_\nu x}-\frac{m^2_\tau}{2E^2_\nu }\right)\right/\left(1+\frac{xM}{2E_\nu}\right), 
		            \\ \\ B=\left.\frac{1}{2} \left[ \left(1-\frac{m^2_\tau}{2ME_\nu x}\right)^2-\frac{m^2_\tau}{E^2_\nu}\right]^{\frac{1}{2}}\right/\left(1+\frac{xM}{2E_\nu}\right).
		  \end{array}
\label{eq:AandB}
\end{equation}

\begin{figure}[t]
    \centering
    \includegraphics[width=9cm,height=9cm]{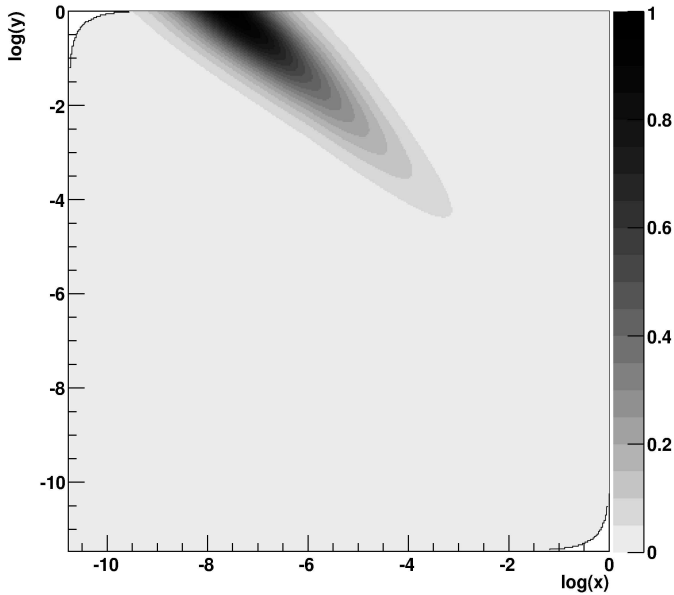}
    \caption{\small{Distribution in the $\log{x}-\log{y}$ plane for the physical region of the $\nu_\tau - N$ charged-current interaction for an initial neutrino energy $E_\nu =
    10^{11}$ GeV.
    We plot the entire available region. The pdf set used here is CTEQ6. The other set does not produce clearly visible differences.}}
    \label{fig:xydistr}
\end{figure}

To the previous region for $y$ given in equations \ref{eq:yrange} and \ref{eq:AandB}, we added the condition:
\begin{equation}
\frac{\left(M^2_\Delta - M^2\right)}{2ME_\nu} \leq y \leq 1-\frac{m_\tau}{E_\nu}
\label{eq:ycond}
\end{equation}
The upper limit is just a safety condition asking for the minimum energy of the produced tau to be $m_\tau$ ($\tau$ produced at rest). The minimum limit chosen here, with
$M_\Delta$ the mass of the $\Delta$ resonance of the nucleon, requires that the interaction always occurs in the deep-inelastic region. This is convenient both to reduce the
computation time and increasing the precision of our $(x,y)$ distributions. For the high energies considered here, this is well justified as deep-inelastic scattering is totally dominant
compared to quasi-elastic scattering or resonance production \cite{hagiwara}.
Fig.\ref{fig:xydistr} shows an example of distribution used for our calculation. It gives the physical region in the log(x)-log(y) plane, as defined above, for CC 
interaction of an incident $\nu_\tau$ of energy $E_\nu=10^{11}$ GeV.

The dependance of such a distribution on the energy is quite easy to sum up. Increasing energy extends the physical region accessible in the $x-y$ plane and shifts the pattern to 
smaller $x$. Of course, we use different distributions for $\nu_\tau$ and $\overline{\nu}_\tau$ scattering.

\subsection{Tau polarization}
\label{subsec:taupolar}

Once we have fulfilled all those requirements we can head to the calculation of tau polarization. The formula for the $\tau^-$ lepton spin polarization quoted from the above
reference reads:

\begin{eqnarray}
\overrightarrow{P}_{\tau^-}=\frac{m_\tau}{ME_\nu}\left[(\overrightarrow{p}+\frac{\overrightarrow{q}}{x})\mathcal{Q}(x,Q^2)\right.
\nonumber\\
\left.\left.+\overrightarrow{p}(1-\frac{\delta_\tau}{x}-y)\overline{\mathcal{Q}}(x,Q^2)\right]\right/\left(P_Q+P_{\overline{Q}}\right)^{-1}
\label{eq:Ptau}
\end{eqnarray}

For the $\tau^+$ lepton produced in an anti-neutrino scattering we have:

\begin{eqnarray}
\overrightarrow{P}_{\tau^+}=-\frac{m_\tau}{ME_\nu}\left[(\overrightarrow{p}+\frac{\overrightarrow{q}}{x})\mathcal{\overline{Q}}(x,Q^2)\right.
\nonumber\\
\left.\left.+\overrightarrow{p}(1-\frac{\delta_\tau}{x}-y){\mathcal{Q}}(x,Q^2)\right]\right/\left(P_Q+P_{\overline{Q}}\right)^{-1}
\label{eq:Ptaubar}
\end{eqnarray}

In these expressions, $\mathcal{Q}$ and $\overline{\mathcal{Q}}$ are defined exactly as above. $\delta_\tau = \frac{m^2_\tau}{2ME_\nu}$ and 
\begin{eqnarray}
&&P_Q=(1-\frac{\delta_\tau}{x}){\mathcal{Q}}
\nonumber\\
&&P_{\overline{Q}}=(1-y)(1-y-\frac{\delta_\tau}{x})\overline{\mathcal{Q}}.
\label{pqs}
\end{eqnarray}
The 3-vectors $\overrightarrow{p}$ and $\overrightarrow{q}$ are the momenta of the proton and of the transfered momentum $q$ respectively. They are both expressed in the
tau rest frame. To obtain the expression of the spin polarization in another frame, one just have to boost the 4-vector $(0,\overrightarrow{P})$.

In the laboratory frame, the spin polarization vector always lies in the scattering plane \cite{hagiwara} so that the spin polarization vector in the tau rest frame can be
written as \begin{equation}\overrightarrow{P} = (P_x, P_y, P_z) = P(\cos{\theta_P},\sin{\theta_P}\cos{\phi_P},0),\end{equation} where $\theta_P$ and $\phi_P$ are
the polar and azimuthal angles of the spin vector in the tau rest frame, respectively, and P denotes the degree of polarization. $P=1$ corresponds to a fully polarized tau,
whereas $P=0$ gives the unpolarized case.

In our case, the useful physical information lies mostly in $P_x/P = \cos{\theta_P}$. The massless particle case would lead to $P_x = \pm 1$, which means fully left-handed
$\tau^-$ or right-handed $\tau^+$. For the non-zero mass $\tau^\pm$ leptons this is not the case anymore.

 In the following, we will be interested in evaluating the value of $P_x/P$, more particularly in the laboratory frame for the UHE range of $10^{-1}$ EeV 
 $\leq E_\nu \leq 10^2$ EeV.
This range is relevant for surface detectors such as the Pierre Auger Observatory and
corresponds to the detection window for the expected GZK neutrinos. For such experiments, the tau polarization represents a large systematic error. This is
easy to understand. Depending on the created $\tau^\pm$ helicity, the decay products have different energy distributions. This influences the
efficiency of trigger and of identification and thus the acceptance of the detector to the emerging tau leptons \cite{pao1,pao2}. From the previous references, we quote a 30\%
difference for the acceptance to $\tau^-$ between the two extreme cases of polarization, i.e. all $-1$ or all $+1$ helicities. It must be noted that $\tau^+$
 with $+1$ helicity gives the same energy distributions than $-1$ $\tau^-$, so that they should lead to the same acceptance. This symmetry is nearly exact, even if small
 corrections remain in the angular distributions of the multihadron final states \cite{tauola}, sensitive to CP parity. But this effect is of course negligible and a full symmetry 
 can be assumed between $\tau^-$ and $\tau^+$.
 
Our study aims at improving our knowledge on the helicity of the produced $\tau^\pm$ leptons in order to constrain the calculation of the acceptance to these
particles and thus to reduce the systematic error due to the lack of information on the polarization.

We will also derive some informations for lower energies, which could be useful for experiments working in this lower range of energy.

\section{Results}
\label{sec:results}

We simulated mono-energetic beams of $\nu_\tau$ and $\overline{\nu}_\tau$ which interact through CC interaction with an isoscalar nucleon at rest, in other words, cosmic
(anti-)neutrinos that interact with the Earth crust.

We are interested in studying different features of the tau polarization both in the tau rest frame and the laboratory frame: the evolution with energy, differences between 
$\tau^-$ and $\tau^+$ leptons and the influence of different pdf sets on our results.

\begin{figure}[t]
    \centering
    \includegraphics[width=9cm,height=9cm]{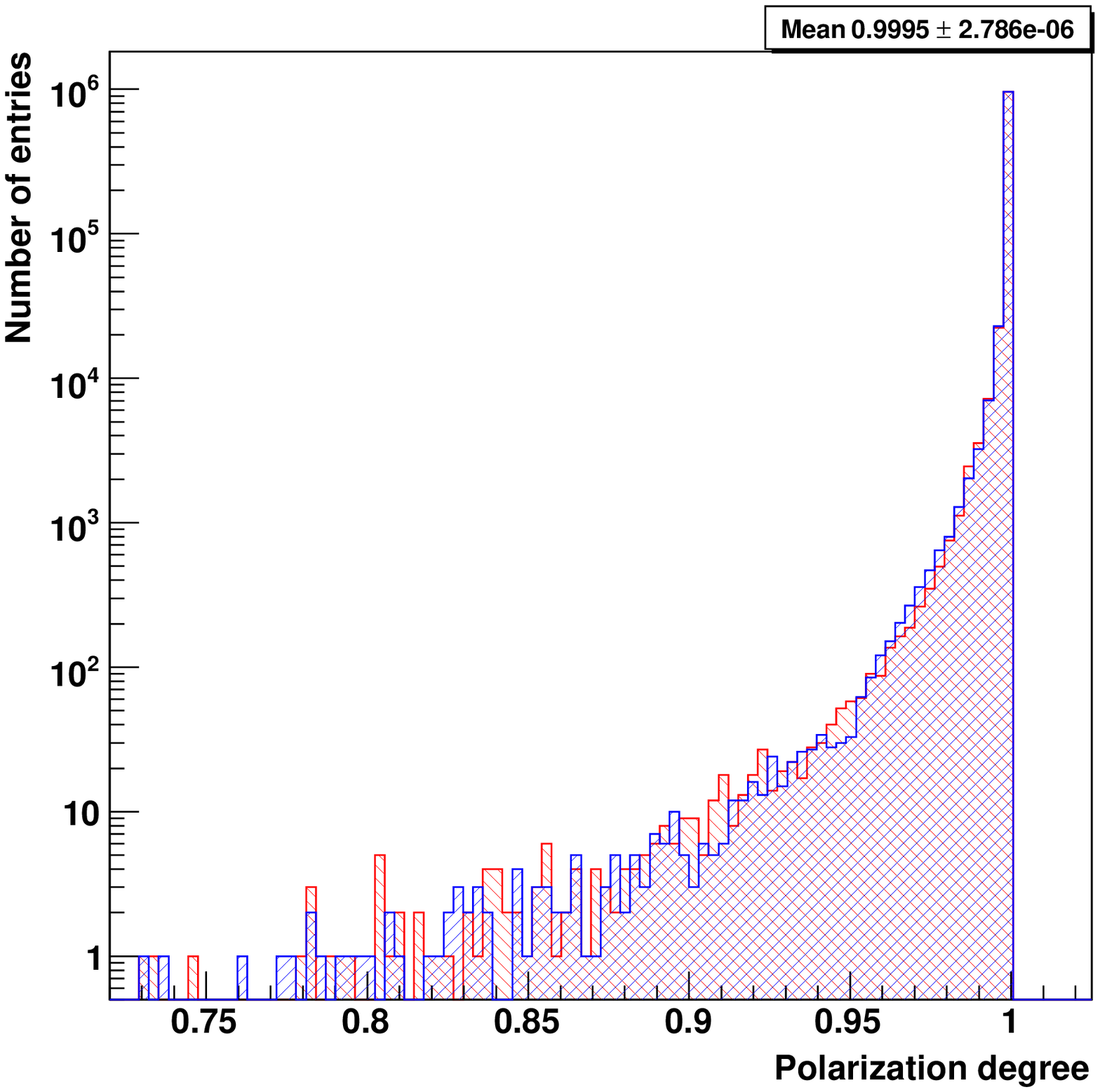}
    \includegraphics[width=9cm,height=9cm]{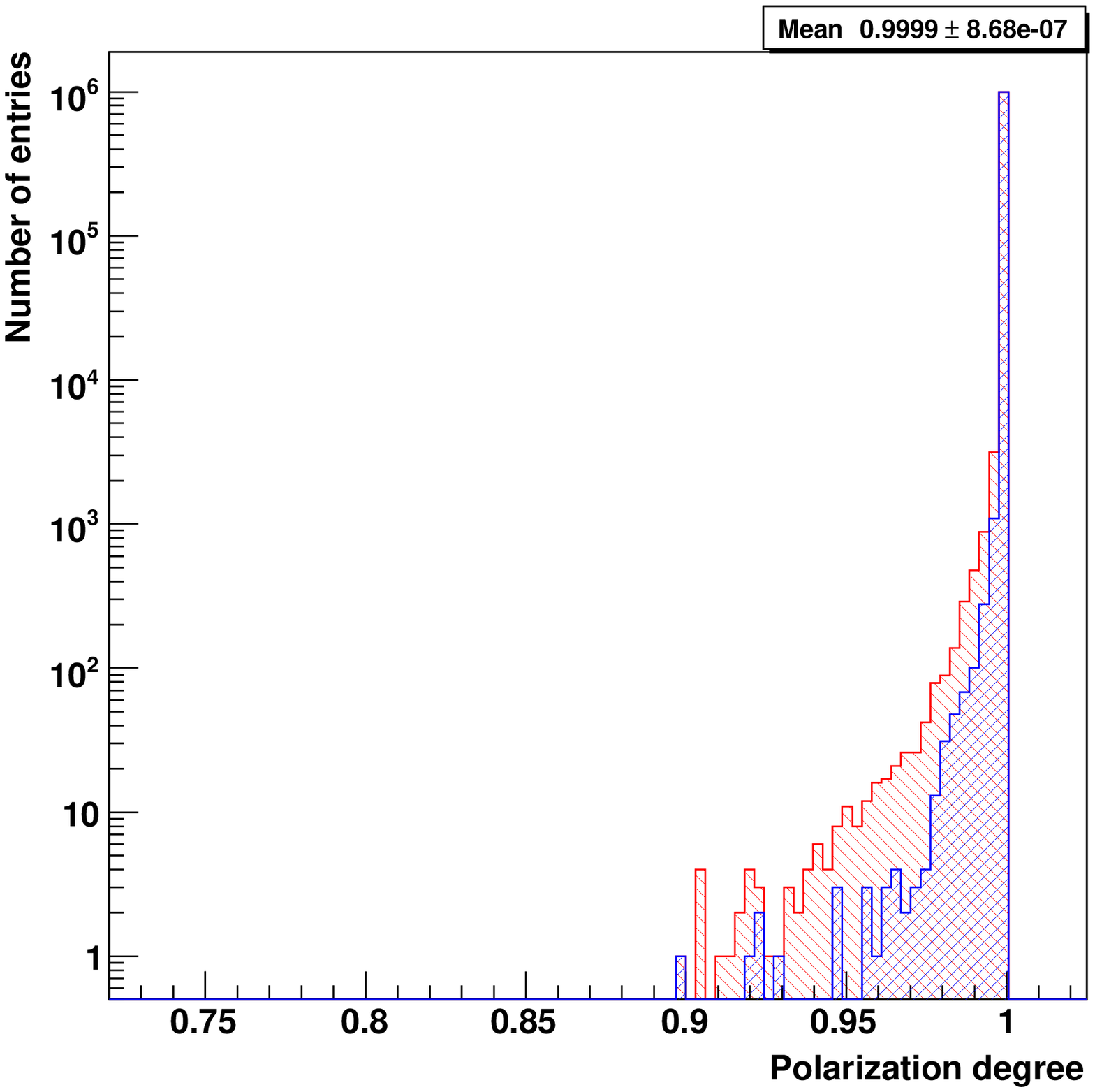}
    \caption{\small{Degree of polarization for $\tau^-$ leptons produced in $\nu_\tau$ CC interaction, from a mono-energetic $\nu_\tau$ beam. The upper figure corresponds to initial neutrinos of energy $E_\nu=10^4$ GeV and the other to $E_\nu=10^{11}$ GeV}}
    \label{fig:P21e11}
\end{figure}

For the discrepancy due to the different pdfs, if we expect a difference, it is at high energy. In figure \ref{fig:P21e11} we show the polarization degree P for $\tau^-$ leptons induced by incident
neutrinos of $10^4$ and $10^{11}$ GeV. This value is given here in the tau rest frame. In red, we plot the distribution obtained using CTEQ6 set and in blue that from
MRST.

The result obtained is in accordance with what we expect: no differences at low energies, contrary to what we observe at higher energies, where
existing informations constraining the values of pdfs have to be extrapolated to regions where we have no way to evaluate the physical evolution. An other information is
the evolution of the polarization degree of the produced $\tau^-$ leptons. The created particles are highly polarized even at low energies. At first order, one could almost
consider the fully polarized case, whatever the energy. For produced $\tau^+$ leptons, the same comments hold.

\begin{figure}[t!]
    \centering
    \includegraphics[width=9cm,height=9cm]{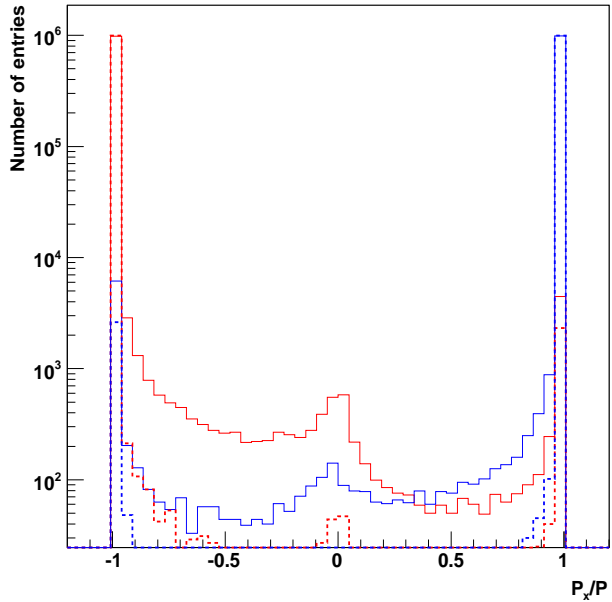}
    \includegraphics[width=9cm,height=9cm]{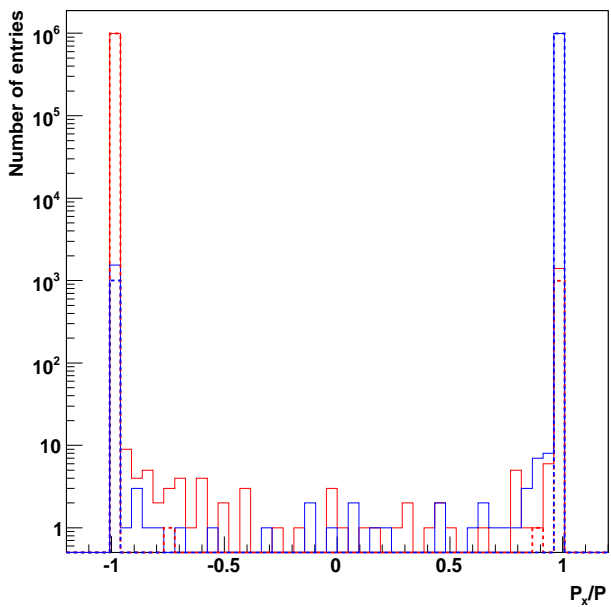}
    \caption{\small{Helicity of $\tau^\pm$ leptons produced in CC interactions. In each graph, we plot in red the distribution for $\tau^-$ leptons and in blue the one for
    $\tau^+$ leptons. In the upper figure, the solid line corresponds to the distribution for $E_\nu = 10^4$ GeV and the dashed line to $E_\nu = 10^5$ GeV. In the bottom figure,
    solid line is for $E_\nu = 10^6$ GeV and dashed line for $E_\nu = 10^7$ GeV}}
    \label{fig:Px1e4-7}
\end{figure}

In figure \ref{fig:Px1e4-7}, we now show the distributions obtained for $P_x/P$, namely the helicity, in the laboratory frame, for $E_\nu=10^4$ GeV to $10^7$ GeV. The red histograms are
for  $\tau^-$ and the blue ones for $\tau^+$. For details on the figures, see the caption. The differences between the tau leptons and their antiparticles appear clearly. 
They should be reduced as energy increases because the parton distributions involved in both interaction ($\nu_\tau N$ and 
$\overline{\nu}_\tau N$) become more and more dominated by sea quarks and tend to the same value. We do not talk here about the difference in the overall sign, which is obvious, but more about the difference in the shape of the
distributions which in consequence should get almost symmetric at very high energy. The $\tau^+$ leptons also tend to be slightly more polarized than the
$\tau^-$ at low energy.

It is obvious that the produced leptons get more and more polarized as the energy increases. For $\tau^-$, the
distribution seems to tend to only two bins, $-1$ and $+1$, the proportion of $-1$ helicity leptons growing with energy. For $\tau^+$ the same conclusions hold by inverting $-1$
and $+1$ helicities.

\begin{figure}[t!]
    \centering
    \includegraphics[width=9cm,height=9cm]{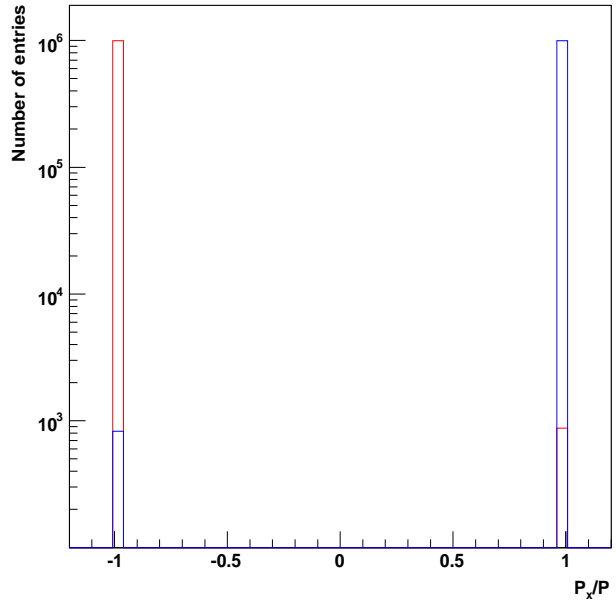}
    \caption{\small{Helicity of $\tau^\pm$ leptons produced in CC interactions. We plot in red the distribution for $\tau^-$ and in blue the one for
    $\tau^+$. We consider incident neutrinos of energy $E_\nu = 10^8$ GeV}}
    \label{fig:Px1e8}
\end{figure}

At higher energies, in the region that we are interested in, this behaviour becomes still more obvious. We show in figure \ref{fig:Px1e8} the same distribution for incident
neutrinos of energy $E_\nu = 10^8$ GeV. We use the same color code than above.

The distribution consists now only in 2 bins in $-1$ and $+1$, which means only left-handed or right-handed particles.
We also computed the distributions for higher energies but it is not useful to show them here as they do not give us much more information than what we learn from the previous
plot. All that we see is that every created
$\tau^-$ (resp. $\tau^+$) are mostly left-handed (resp. right-handed) with a ratio of $1000:1$ in favour of $-1$ helicity (resp. $+1$). At $E_\nu=10^{11}$ GeV the mean helicity is
$-0.9983$ (resp. $0.9982$).

This result is exactly what is needed in order to address the problem of the systematic uncertainties due to the tau polarization. In view of the values shown above, it is totally
fair to consider that all produced $\tau^-$ have $-1$ helicity and that all $\tau^+$ are produced with a $+1$ helicity.

All the previous results were obtained by using CTEQ6 pdf set. We also computed the same distributions using MRST to check the influence of the set used. At highest
energies, we obtain a slightly larger proportion of $+1$ and $-1$ helicity for $\tau^-$ and $\tau^+$ respectively but this is only a second order effect. The two different sets used
in this work represent well the range of the distribution functions present throughout the litterature and hence the present result can reasonably be assumed to be independant
from the pdf choice.

\section{Conclusion}
\label{sec:conclusion}

In this paper we have studied the polarization vector of $\tau^\pm$ leptons produced in $\nu_\tau$ or $\overline{\nu}_\tau$ deep-inelastic interaction with the Earth crust, in
order to address the problem of the systematic uncertainties in the frame of large surface detectors such as the Pierre Auger Observatory.

Our main result allows to choose unambiguously between $-1$ or $+1$ helicities, for both $\tau^-$ and $\tau^+$ leptons, when studying their propagation or decay.
This result has been shown to be only lightly dependant on the parton distribution function set used. The ratio $-1/+1$ particles may be slightly affected, but this is just a second order
effect that cannot impact deeply on any study following our conclusions.

The last point, along with the fact that $-1$ helicity $\tau^-$ and $+1$ helicity $\tau^+$ lead to the same energy
distributions among their decay products, allows to choose between the two extreme cases of acceptance to up-coming $\tau^\pm$ leptons and so to eliminate the $\tau$ polarization as a source of
systematic uncertainties.

This conclusion holds when considering usual pdf sets but the study could be repeated using exotic models for the evolution of the pdfs, to quantify the influence of new physics.

Below $10^8$ GeV, even if the $-1$ helicity case is prevailing, the picture is not so contrasted, so that for studies concerning this lower energy range, extra care should be
taken in evaluating the tau polarization distribution and its
impact on their particular problem.

\section*{Acknowledgments}
The author would like to thank his colleagues of the Pierre Auger Collaboration for useful discussions and suggestions. This work was supported by a PhD fellowship of the French
Ministry for Education and Research.

\end{document}